\documentclass[sigconf, camera-ready]{acmart}

\AtBeginDocument{%
  \providecommand\BibTeX{{%
    \normalfont B\kern-0.25em{\scshape i\kern-0.25em b}\kern-0.5em\TeX}}
    \setlength\abovedisplayskip{0.2pt}
    \setlength\belowdisplayskip{0.4pt}}

\setcopyright{acmcopyright}
\copyrightyear{2020}
\acmYear{2020}

\definecolor{background}{HTML}{EEEEEE}
\definecolor{delim}{RGB}{20,105,176}
\definecolor{mygreen}{rgb}{0.13, 0.55, 0.13}
\definecolor{mygold}{rgb}{1.0, 0.75, 0.0}
\definecolor{myred}{rgb}{0.6, 0.0, 0.0}
\definecolor{myblue}{rgb}{0.25, 0.4, 0.96}
\definecolor{mypurple}{rgb}{0.56, 0.0, 1.0}
\definecolor{deepskyblue}{rgb}{0.39, 0.58, 0.93}

\usepackage[compact]{titlesec}
\titlespacing{\section}{0pt}{1ex}{0ex}
\titlespacing{\subsection}{0pt}{1ex}{0ex}
\titlespacing{\subsubsection}{0pt}{1ex}{0ex}

\newcommand{\specialcell}[2][l]{%
  \begin{tabular}[#1]{@{}l@{}}#2\end{tabular}}

\begin{document}

\title{COVID-19 Knowledge Graph: Accelerating Information Retrieval and Discovery for Scientific Literature}


\author{Colby Wise, Vassilis~N.~Ioannidis, Miguel~Romero~Calvo, Xiang~Song, George~Price, Ninad~Kulkarni, Ryan~Brand, Parminder~Bhatia, George~Karypis}
\affiliation{%
  \institution{Amazon Web Services AI}
}
\email{colbywi, ivasilei, miguelrc, xiangsx, gwprice, ninadkul, brandry, parmib, gkarypis @amazon.com}

\renewcommand{\shortauthors}{Wise and Ioannidis, et al.}

\begin{abstract}
The coronavirus disease (COVID-19) has claimed the lives of over 350,000 people and infected more than 6 million people worldwide. Several search engines have surfaced to provide researchers with additional tools to find and retrieve information from the rapidly growing corpora on COVID-19. These engines lack extraction and visualization tools necessary to retrieve and interpret complex relations inherent to scientific literature. Moreover, because these engines mainly rely upon semantic information, their ability to capture complex global relationships across documents is limited, which reduces the quality of similarity-based article recommendations for users. In this work, we present the COVID-19 Knowledge Graph (CKG), a heterogeneous graph for extracting and visualizing complex relationships between COVID-19 scientific articles. The CKG combines semantic information with document topological information for the application of similar document retrieval. The CKG is constructed using the latent schema of the data, and then enriched with biomedical entity information extracted from the unstructured text of articles using scalable AWS technologies to form relations in the graph. Finally, we propose a document similarity engine that leverages low-dimensional graph embeddings from the CKG with semantic embeddings for similar article retrieval. Analysis demonstrates the quality of relationships in the CKG and shows that it can be used to uncover meaningful information in COVID-19 scientific articles. The CKG helps power \url{www.cord19.aws} and is publicly available. 
\end{abstract}



\maketitle

\section{Introduction}
\label{intro}
The onset of the novel SARS-CoV-2 virus has emphasized the need to accumulate insights from large volumes of information. Thousands of new scientific articles on the virus are being published weekly, leading to a rapid increase in the cumulative knowledge about the coronavirus disease (COVID-19). COVID-19 has heightened the need for tools that enable researchers to search vast scientific corpora to find specific information, visualize connections across the data, and discover related information in the data. 

Several COVID-19 dedicated search engines have come online to address the need for information retrieval of scientific literature on the disease. Search engines like Sketch Engine COVID-19, Sinequa COVID-19 Intelligent Search, Microsoft’s CORD19 Search, and Amazon’s CORD19 Search use a variety of methodologies such as keyword search, natural language queries, semantic relevancy, and knowledge graphs. 
However, these engines return thousands of search results that overlook inherent relationships between scientific articles, such as subject topic and citations, and do not provide tools to visualize relationships, which is beneficial for knowledge discovery. In this paper, we construct the COVID-19 knowledge Graph (CKG) by extracting rich features and relationships of COVID-19 related scientific articles and develop a document similarity engine that combines both semantic and relationship information from CKG.

Knowledge graphs (KGs) are structural representations of relations between real-world entities where relations are defined as triplets containing a head entity, a tail entity, and the relation type connecting them. KG based information retrieval has shown great success in the past decades~\cite{kim1990model, dalton2014entity}.
We construct the  CKG  using the  CORD19 Open Research Dataset of scholarly articles~\cite{wang2020cord}. Scientific articles, publication authors, author institutional affiliations, and citations form key relationships in the graph. Further, we extract  biomedical entity relationships and highly abstracted topics from the unstructured text of articles using Amazon Comprehend Medical service and  train a topic model on the corpus. By applying data normalization technologies we eliminate duplicate entities and noisy linkages. The resulting KG contains 336,887 entities and 3,332,151 relations. The CKG has been made publicly available to researchers with rapid “one-click” cloud deployment templates.\footnote{https://aws.amazon.com/cn/covid-19-data-lake/}
We introduce a document similarity engine that leverages both the semantic information of articles and the topological information from the CKG to accelerate COVID-19 related information retrieval and discovery. We employ SciBERT~\cite{beltagy2019scibert}, a pretrained NLP model,  to generate semantic embeddings for each article. Meanwhile, we utilize knowledge graph embedding (KGE)~\cite{wang2017knowledge,zheng2020dgl} and graph neural network~\cite{schlichtkrull2018modeling} technologies to generate embeddings for entities and relations of the CKG. Finally, by combining judiciously  the semantic embeddings and graph embeddings we use  the similarity engine to propose top-k similar articles. The CKG and similarity engine are new additions to \url{www.CORD19.aws}, a website using machine learning to help researchers search thousands of COVID-19 related scientific articles using natural language question queries that has seen over 15 million queries across more than 70 countries. The CKG adds a graph-traversal ranking feature to search and the similarity engine powers the similarity-based recommended article system. To further demonstrate the quality of the CKG, we conduct a series of experiments analyzing the relations that form the core pillars of the graph. We first evaluate the ability of our methodology to capture the topic information in the text, and show that extracted topics align well with the subjects of scientific journals. We also perform link prediction analysis by extracting graph embeddings that validates the quality of the relations in the graph and demonstrates that we capture important  topological information from the CKG. Our analysis shows that the semantic embeddings and graph embeddings learn useful information and improve our ability to quantify similarity between articles. Lastly, several motivating examples show that querying the CKG can extract actionable insights from scientific articles. To summarize, our contribution is fourfold:
~\begin{itemize}
\item[\textbf{C1}] We construct a scientific KG, named COVID-19 Knowledge Graph (CKG), by judiciously combining the inherent schema information from COVID-19 scientific articles as well as the extracted biomedical entity relationships and topic information.

\item[\textbf{C2}] We conduct several data normalization methodologies to curate the CKG and demonstrate its information retrieval, visualization and discovery capabilities. The CKG is publicly available through \url{https://aws.amazon.com/cn/covid-19-data-lake/}. 

\item[\textbf{C3}] We present a novel similarity-based document retrieval system that combines semantic article information with document topological information learned from the CKG and show that it reliably improves the quality of user-suggested documents.

\item[\textbf{C4}] The similarity engine and the CKG have been integrated into a public available search service for COVID-19 through \url{www.CORD19.aws} to power the similarity-based article recommendation system and to provide a graph-traversal ranking feature.
\end{itemize}

\section{CKG Construction \& Curation}
\label{Structure of Graph}

\begin{figure}[H]
    \captionsetup{format=plain}
    \centering
    \includegraphics[width=1\linewidth]{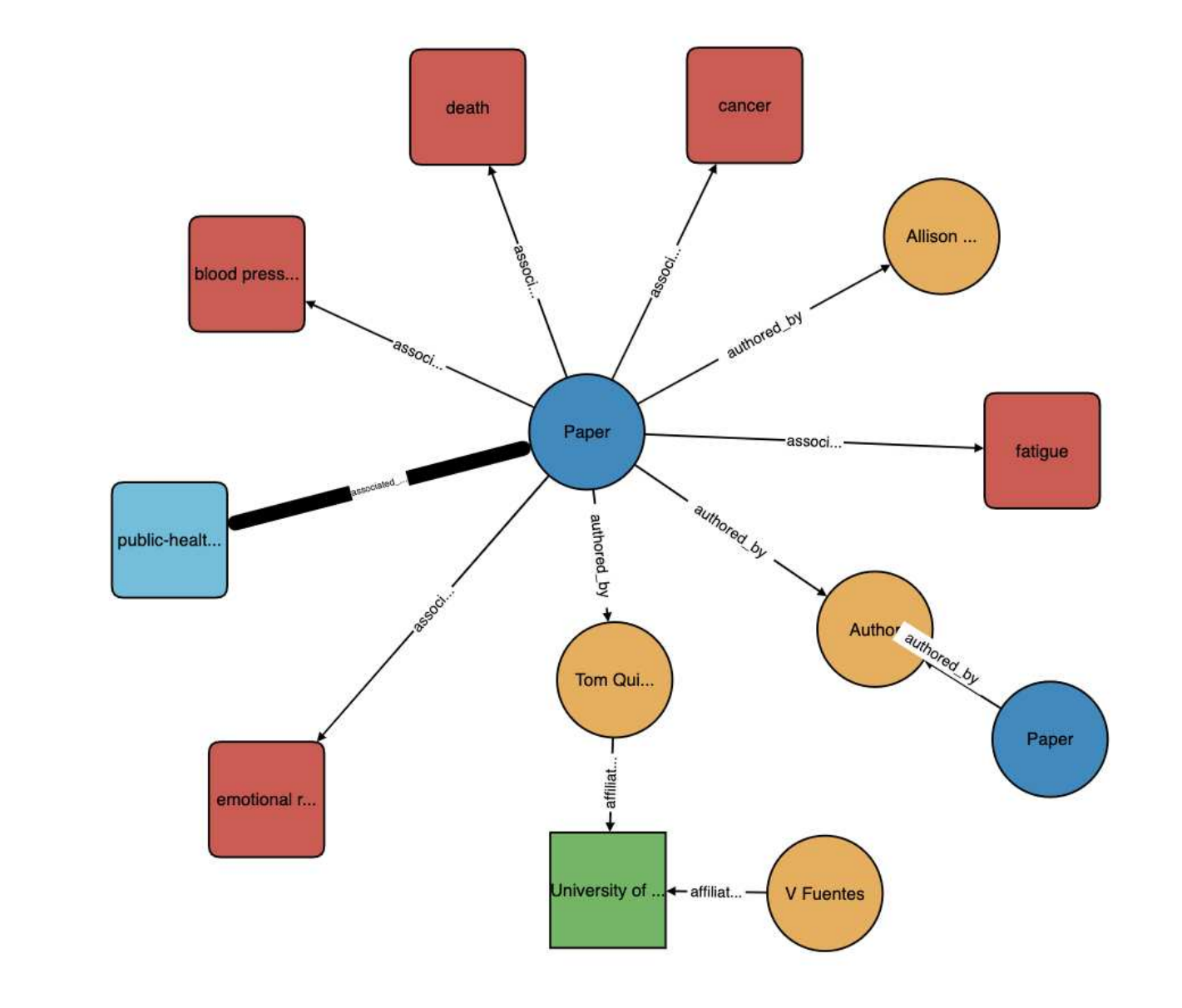}
    \caption{Visualization of CKG. Paper entities (\textcolor{myblue}{blue}) connect to Concepts (\textcolor{myred}{red}), topics (\textcolor{deepskyblue}{light blue}), and authors (\textcolor{mygold}{gold}) through directed relations. Authors connect to institutions (\textcolor{mygreen}{green}).}
    \label{fig:CKG}
\end{figure}

CKG is a directed property graph where entities and relations have associated attributes (\textit{properties}) and direction (\textit{directed}). Figure~\ref{fig:CKG} illustrates the directed property graph structure for a small subgraph of CKG. In this section we describe the dataset used to construct the CKG, define the entity and relation types, detail CKG curation methods, provide summary statistics describing the graph, and detail the cloud infrastructure that drives CKG scalability.

\vspace{-0.4cm}
\vspace{-.5cm}

\subsection{The CORD-19 Dataset}
\label{CORD-19 DataSet}

COVID-19 Open Research Dataset (CORD-19) is a dynamic, growing repository of scientific full text articles on COVID-19 and related coronaviruses created by the Allen Institute for AI (AI2)~\cite{wang2020cord}. The data is made available via Kaggle with weekly updates as part of the on-going CORD-19 Research Competition~\cite{kagcord}. 

As of 06-01-2020, the CORD-19 dataset consists of over 60,000 full text. Rich metadata is provided as part of the dataset e.g. article authors.  The data is sourced from several channels such as PubMed, bioArxiv, and medRxiv. The dataset is multidisciplinary with articles covering virology, translational medicine, epidemiology, computer science and more. CORD-19 grows constantly and AI2 is working with the wider research community to normalize and improve the quality of the data. 

\subsection{Entity Types}

The CKG contains five types of entities corresponding to papers, authors, institutions, concepts, and topics as summarized in Table \ref{tab:graphStats}. Information on what these entities represent, their attributes, and how they are created follows.

\paragraph{Paper Entities.} Representation of scientific articles. Attributes include title, publication date, journal, and Digital Object Identifier (DOI) link as available in the CORD-19 Dataset from AI2.

\begin{table}[H]
\caption{COVID-19 Knowledge Graph entity and relations.}
\small
\begin{tabular}{lrlr}\toprule
\textbf{Entity Type} & \textbf{Count} & \textbf{Relation Type} & \textbf{Count}   \\ \midrule
Papers         & 42,220  & authored\_by       & 240,624  \\ 
Authors        & 162,928 & affiliated\_with & 121,257  \\ 
Institutions   & 21,979  & associated\_concept      & 2,739,665 \\ 
Concepts       & 109,750 & associated\_topic.        & 95,659   \\ 
Topics         & 10      & cites        & 134,945  \\ \midrule
\textbf{Total} & 336,887 &                    & 3,332,151 \\ \bottomrule
\end{tabular}
\centering
\label{tab:graphStats}
\end{table}

\paragraph{Author Entities.} Representation of the paper authors. Attributes include the first, middle, and last names.

\paragraph{Institution Entities.} Institution affiliations for authors. Attributes include institution name, country, and city. 

\paragraph{Concept Entities.} Comprehend Medical (CM) Detect Entities V2 is an Amazon Web Service that uses natural language processing (NLP) and machine learning for medical language entity recognition and relationship extraction~\cite{cm}. CM classifies extracted entities into entity types: Ibuprofen (entity) belongs to the Medications category (entity type). We leverage CM to extract biomedical entities from the scientific articles. Specifically, given the example text "Abdominal ultrasound noted acute appendicitis, recommend appendectomy followed by several series of broad spectrum antibiotics," CM extracts \textit{Abdominal} (Anatomy), \textit{ultrasound} (Test Treatment Procedure), \textit{acute appendicitis} (Medical Condition), \textit{appendectomy} (Test Treatment Procedure), and \textit{antibiotics} (Medication) as recognized entities along with model confidence scores. Entity names e.g. \textit{acute appendicitis}, form concept entities in the CKG while entity category and model confidence score are the entities' attributes. 

\paragraph{Topic Entities.} We use an extension of Latent Dirichlet Allocation (LDA) \cite{blei2003latent} termed Z-LDA \cite{andrzejewski2009latent}, trained using the title, abstract and body text from each paper. Labels are generated with the help of medical professionals to eliminate, merge, and form 10 topics which serve as the basis for topic entities in the CKG: Vaccines/Immunology, Genomics, Public Health Policies, Epidemiology, Clinical Treatment, Virology, Influenza, Healthcare Industry, Lab Trials (human) and Pulmonary infections. Re-modeling and manually labeling a topic model is inefficient, therefore we train a multi-label classifer~\cite{read2011classifier} using the original topic model labels and a training split from 59k total documents. The resulting classifier achieves an average F$_{1}$ score of 91.92 with on average 2.37 labels per document.

\subsection{Relation Types}

Relations in the CKG are directed and summarized in Table \ref{tab:graphStats}. Here we defined all relation types.

\paragraph{authored\_by.} This relation connects paper entities with author entities and indicates that authorship relation.

\paragraph{affiliated\_with.} This relation connects author entities with institution entities and indicates that affiliated relation.

\paragraph{associated\_concept.} This relation connects paper entities with concept entities and indicates that associated relation. These relation have the CM model confidence score as an attribute.

\paragraph{associated\_topic.} This relation connects paper entities with topic entities and indicates that associated relation. These relation have the Z-LDA prediction score as an attribute.

\paragraph{cites.} This relation connects paper entities with paper references indicating a citation relation.

\subsection{CKG Curation} 
\label{CKG Curation}

\subsubsection{Concept Normalization}\hfill\\
We use thresholding on the confidence scores as a de-noising step by requiring an entity's confidence scores to exceed a $0.5\%$ threshold that is determined through empirical experimentation. {We explored a parameter range of $0.4\%$-$0.6\%$ in $0.1$ increments.} Thresholding comes at the expense of entity coverage: higher confidence threshold increases the likelihood of papers with no or few extracted entities. Next, we lemmatize concept entity names as a form of normalization using  SciSpacy~\cite{neumann2019scispacy}. SciSpacy is built upon the robust SpaCy NLP library~\cite{spacy}, but is trained on biomedical texts similar to those in the CORD19 dataset. We experimentally found SciSpacy to provide target results for limited string lemmatization test cases. Moreover, we keep a running distribution of concept appearances across the dataset. A concept may appear in $N$ papers, where $N$ is the total number of papers in the dataset. We prune concepts that occur in less than \text{0.0001\%}. Concepts that appear in greater than \text{50\%} are flagged for manual qualitative assessment of information value. The main downside of this approach is scalability and in future work we plan to systematize and extend this process using domain-specific specialized ontology standardization tools like Comprehend Medical RxNorm~\cite{cmmedic}.

\subsubsection{Author Normalization}\hfill\\
Author names in the CORD-19 dataset require judicious processing. Oftentimes, paper authors have incomplete information such as missing “first name” or high name variation between different academic journals. Additionally, author citations often follow an abbreviated format using “first initial, last name”. We utilize a hybrid approach similar to \cite{ammar2018construction} involving normalization and linking. When linking authors, we normalize author names via lower casing, removing punctuation, and merging “first, middle, last name”.

\subsubsection{Citation Linking}\hfill\\
We also normalize the author information in the cited papers and match the normalized author names. This allows us to link papers based on citations. 
 
We require that both normalized author information and article title information match exactly. From here, we include citation links for papers referenced within the CORD-19 dataset and find ~$43\%$ of papers cite another paper available in the CORD-19. 

\begin{figure}[t]
    \captionsetup{format=plain}
    \centering
    \includegraphics[width=1\linewidth]{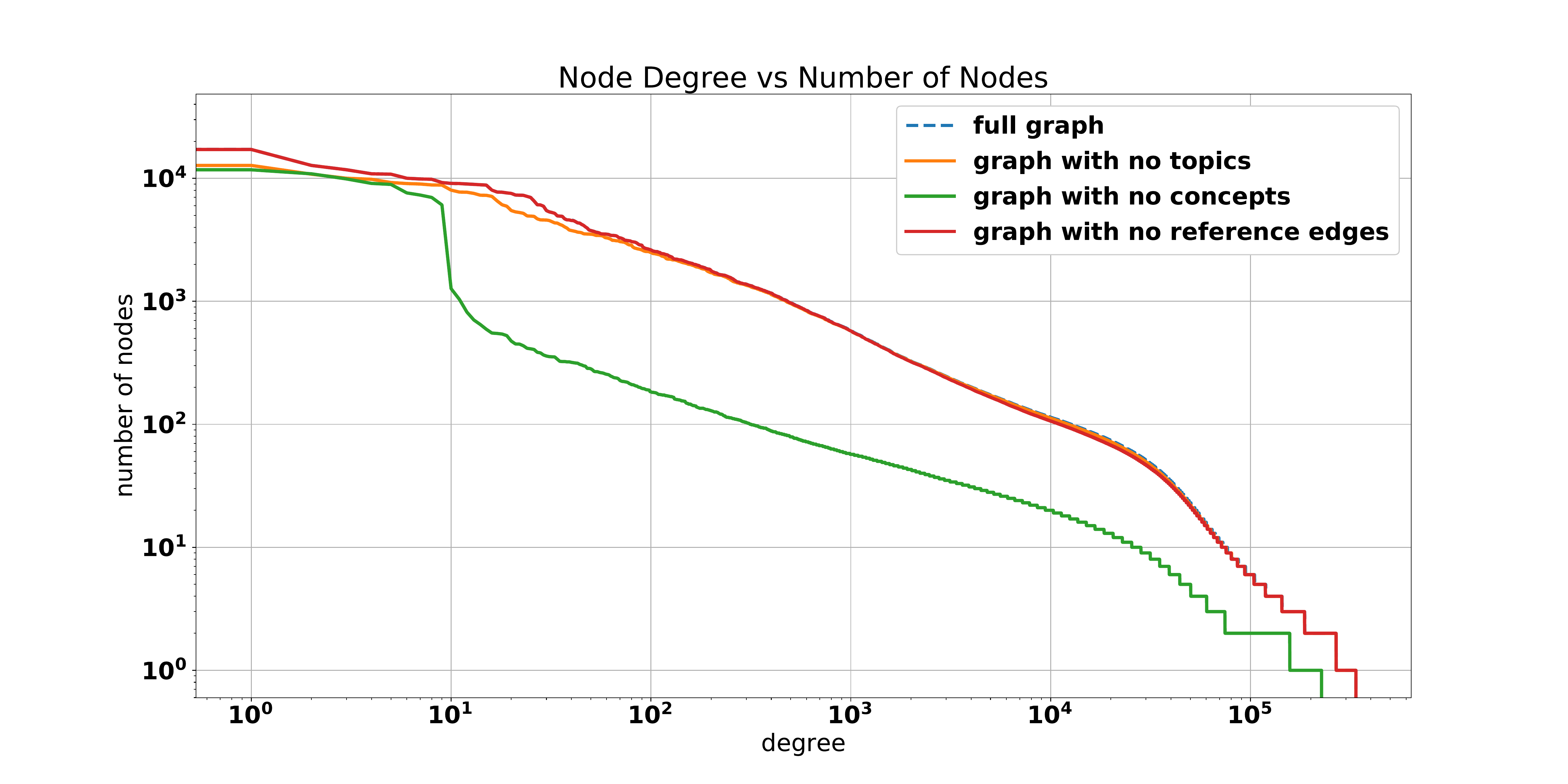}
    \caption{Degree distribution of CKG for various sub-graphs: shows degree change of CKG with \textcolor{mygreen}{concept} relations removed; \textcolor{myred}{citation} relations removed; \textcolor{orange}{topic} relations removed.}
    \label{fig:degree}
\end{figure}

\subsection{Graph Statistics} 
Table  \ref{tab:graphStats} provides counts of all entity and relation types. The $\sim$42k paper entities have on average 2.3 outgoing topic relations, 64.9 outgoing relations to concepts, 5.7 outgoing relations to authors 
\vspace{-.2cm}
and 3.2 outgoing citation relations. Furthermore, $\sim$29k paper entities have at least one outgoing citation relation to another paper, $\sim$18k have at least one incoming citation relation from another paper, $\sim$14.6k have at least one incoming and outgoing citation relation, and $\sim$9.7k have neither an incoming nor outgoing citation relation. The 163k author entities have on average $0.75$ outgoing relations to institutions indicating not all authors have institution information in the data. When considering an undirected version of the graph, there are $109$ connected components with the diameter of the largest connected component (CC) equaling $12$ entities that indicates one large CC contains 99\% of relations and entities, while the diameter ($12$) indicates the CKG is dense. 
Figure \ref{fig:degree} shows the undirected degree distribution plot of several sub-graphs of the CKG. We observe that the greatest change in degree distribution comes from the sub-graph without concept relations, exemplifying that concepts form key links in the graph.

\subsection{Infrastructure}\label{infrastructure} 

We use Amazon Neptune, a fully-managed graph database optimized for storage and navigation scaling to billions of relationships. Neptune supports property graphs and the query languages like Apache TinkerPop Gremlin and SPARQL. Neptune's Bulk Loading~\cite{bulkload} feature helps reduce data ingestion time from several hours (sequential loading) to minutes for ~$330k$ entities and ~$3.3M$ relations using a db.r5.4xlarge (8 cores, 16 vcpu, 128 Gb Memory, 3500 Mbps storage bandwidth) Amazon Elastic Compute Cloud (EC2) instance. By utilizing~\cite{covidlake} users can find the exported Neptune graph data, Amazon CloudFormation~\cite{cloudform} templates for one-click recreation and deployment of the CKG, and the structured entities and relation files as comma-separated values (CSV) files. We use Tom Sawyer Graph Database browser for visualizations~\cite{tomsa}.

\begin{figure}[t]
    \captionsetup{format=plain}
    \centering
    \includegraphics[width=1\linewidth]{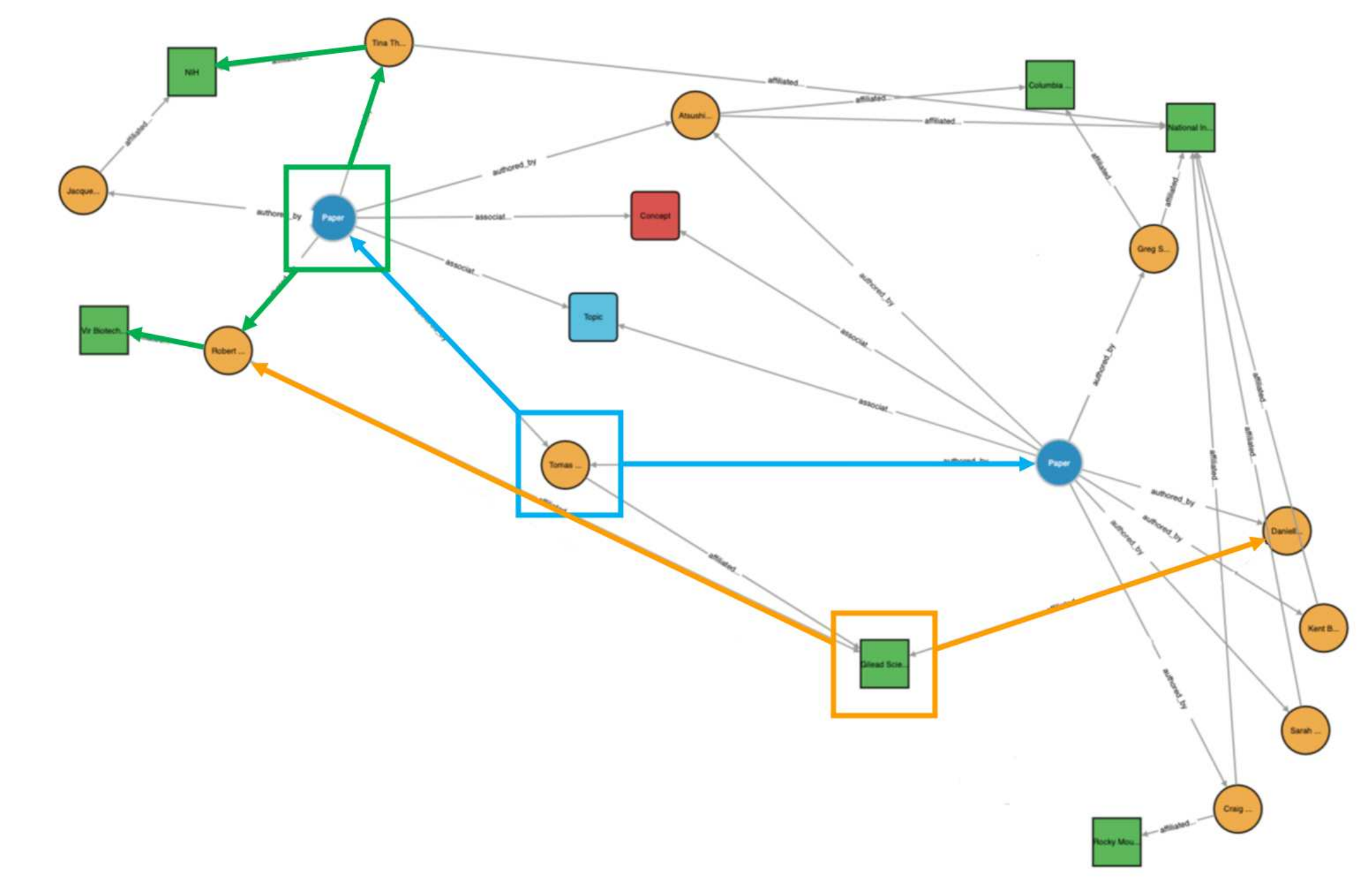}
    \caption{
        Query 1: author research leaders [\textcolor{myblue}{blue box}] ii) institutional leaders [\textcolor{mygold}{gold box}] iii) institution collaborations [\textcolor{mygreen}{green}]  
    }
    \label{fig:application1}
\end{figure}


\section{Using CKG for Information Retrieval}
\label{Example Applications}
In this section we show the CKG uncovers intricate relationships in CORD-19 scientific articles that can aid the research and policy decision processes. 

\begin{itemize}
\item \textbf{Query 1}: What authors and institutions are publishing research pertaining to the drug \textit{remdesivir} and \textit{human lab trial}?
\end{itemize} 

COVID-19 has highlighted the difficulty of health and public policy decision making during pandemics. The above question is motivated by the scenario where policy makers are interested in forming a task force of leading authors and institutions on a rapidly evolving area of research such as a drug treatments for COVID-19. Remdesivir is an investigational nucleotide analog drug currently in FDA clinical trials by Gilead Sciences~\cite{gil}. A CKG user can structure a query identifying articles with \textit{remdesivir} concept and \textit{lab trials (human)} topics form connections. Paper to concept and topic relations form "one-hop" relations. From here we find paper to author relations via another "one-hop" operation and subsequently, author to institution relations via a second "one-hop" (two-hops total) operation. Figure \ref{fig:application1} visually depicts this query process using a small subset of the graph. The author entity, surround by a blue box, is connected to two papers discussing both \textit{remdesivir} and \textit{lab trials} (blue arrows). This author can be viewed as research leader for this query. Similarly, the institutional research leader of this sub-graph is the vertex surrounded in gold box, connected to multiple authors who have published articles matching this query. Lastly, the CKG also helps to uncover multiple-organization collaborations depicted by the vertex surrounded by green box and arrows.

\begin{itemize}
\item \textbf{Query 2}: What papers discussing \textit{COVID-19 risk factors} are most often cited by researchers within the CORD-19 dataset?
\end{itemize}

Researchers can query the CKG to return scientific articles related to specific COVID-19 risk factors such as asthma, heart disease, and respiratory malfunction. The query returns articles with related risk factors. Next, the citation network is leveraged to rank articles by citation counts within the data set. Table \ref{tab:riskfactors} shows the top three results for this query and the respective citations.

\begin{table}[H]
\centering
\footnotesize
    \caption{Graph query results.}
    \begin{tabular}{llc}\toprule
    \textbf{CORD\_UID}  & \textbf{Title} &\textbf{Cited By} \\\midrule
   \textbf{\textit{grw5s2pf}} & The Molecular Biology of Coronaviruses & 498   \\ \hline
   \textbf{\textit{m1jbpo5l}} &\specialcell{Bocavirus and Acute Wheezing\\ in Children}   & 152  \\ \hline
   \textbf{\textit{vnn4135b}} & \specialcell{A Diverse Group of Previously Unrecognized\\  Human Rhinoviruses Are Common \\ Causes of Respiratory Illnesses in Infants}   & 68  \\\hline
    \end{tabular}
    \centering
    \label{tab:riskfactors}
\end{table}

\section{Using CKG for article recommendations}
\label{Similar Paper Engine}

In this section we combine article semantic information with CKG topological information to quantify similarity between articles and construct a similarity-based recommendation system.

\subsection{Leveraging Embeddings}\label{Leveraging Embeddings}

\subsubsection{Semantic Embeddings}\hfill\\
 In order to capture semantic information across the CORD-19 scientific articles we leverage SciBERT \cite{beltagy2019scibert} that has shown strong transfer learning performance on a wide variety of NLP tasks \cite{cer2018universal}. Specifically, our goal is to represent CORD-19 scientific articles as dense document embeddings. 

Sentence Transformer library creates sentence level embeddings from the plain text articles~\cite{reimers2019sentence}. We tokenize the title, abstract and body text into sentences and then using SciBert to create three embedding matrices representing sentences from component of the article. Next, we  average each metric to compute three dense vectors. Finally, a single dense document embedding is obtained by averaging the vectors.

Table \ref{tab:scibertSimilarity} shows the average pairwise cosine similarity of the semantic embeddings constructed from the title, abstract, and body. The cosine similarity matrix among paper pairs is averaged to obtain average similarity for each text portion. We observe the average similarity of scientific articles and availability in the dataset differ based on the article text portion used, noting titles on average have lower similarity and have the highest dataset coverage compared to abstracts or body text. The lower coverage of abstracts drove our decision to combine body and title text with abstracts. 

\begin{table}[H]
\caption{Average cosine distance and percent of dataset coverage using SciBERT embeddings.}
\label{tab:scibertSimilarity}
\begin{tabular}{lcc}\toprule
\textbf{Text Type} & \textbf{Cosine Similarity$_{avg}$} & \textbf{Data Coverage} \\ \midrule
title$_{t}$        & .266   & 97.7\%   \\ 
abstract$_{a}$     & .139   & 84.9\%   \\ 
body$_{b}$         & .092   & 98.6\%   \\ \midrule
$combined$        & .131   & 99.8\%   \\ \bottomrule 
\end{tabular}
\centering
\end{table}

\subsubsection{Knowledge Graph Embeddings: TransE}\hfill\\
\label{TransE}
We leverage knowledge graph embedding (\textit{termed KGE}) methodology to encode entities and relations (relations) in the COVID-19 Knowledge Graph as $d$-dimensional vector embeddings. The embeddings associated with the entities and relations of the graph are generated by a specific KGE algorithm TransE~\cite{bordes2013translating} that satisfy a predetermined mathematical model. We can use these embeddings for downstream tasks such as paper recommendation \cite{zhang2019can}.  In particular, papers with high similarity in embedding space will be highly correlated.

The knowledge graph $G$ is composed of entities and relations such that $G=(V, E)$, where $V$ represents graph entities and $E$ represents the set of relations connecting entities. A specific instance of a relation is represented as a triplet $(h, r, t)$, in which $h$ is the head entity, $r$ the type of the relation, and $t$ the tail entity. Given a set of triplets $T$ in the above format, TransE learns a low-dimensional vector for each entity and relationship where $h + r \sim t$ by minimizing a margin-based objective function over the training set using stochastic gradient descent 
\begin{align}
\min\sum_{\mathbf{h},\mathbf{r},\mathbf{t} \in \mathbb{D}^+ \cup \mathbb{D}^-}\log(1+\exp(-y \times f( \mathbf{h}, \mathbf{r}, \mathbf{t} ) ) )
\label{eq:kgeloss2}
\end{align}
where $f(\boldsymbol{h}, \boldsymbol{r}, \boldsymbol{t}) = \gamma -\left \| \boldsymbol{h} + \boldsymbol{r} -\boldsymbol{t} \right \|_{2}$ is the scoring function;  $\mathbf{h}$, $\mathbf{r}$, $\mathbf{t}$ are the embedding of the head entity $h$, relation ${r}$ and the tail entity $t$, and $\gamma$ is a predefined constant. Here $\mathbb{D}^+$ and $\mathbb{D}^-$ represent the positive and negative sets of triplets respectively, and $y=1$ if the triplet corresponds to a positive example and $-1$ otherwise. Negative triplets are corrupted versions of the extant (positive) triplets defined by the KG, in which either the head or the tail entity have been randomly swapped for another entity in $V$.

We leverage the Deep Graph Library Knowledge Embedding library (DGL-KE)~\cite{zheng2020dgl}, a high performance package for learning large-scale KGE, to train the aforementioned KGE model. By supplying the model with both the entities and relation triplets as described in table~\ref{tab:graphStats}, we generate vector embeddings for each paper.

\subsubsection{Relational Graph Convolutional Network}\hfill\\
KGE models generate embeddings solely by taking into account the structure of the graph. Nevertheless, the learned semantic embeddings can be used as relation features for learning paper relation embeddings. In this section we present an experiment extending the KGE methodology by directly incorporating semantic information to learn paper embeddings that directly capture semantic and topological information. While KGE models do not directly exploit relation features graph convolutional networks can exploit such relation features and possibly obtain richer embeddings~\cite{kipf2016semi}. For this purpose, we apply a relational graph convolutional network (\textit{termed RGCN}) model~\cite{schlichtkrull2018modeling} to learn the relation embeddings exploiting both paper semantic features as well as the graph structure.

An RGCN model is comprised by a sequence of RGCN layers. The output of the $l$th RGCN layer for relation $n$ is a nonlinear combination of the hidden representations of neighboring entities weighted based on the relation type. The relation features are the input of the first layer in the model, which are the semantic paper embeddings. For relation types without features we use an embedding layer that takes as input an one-hot encoding of the relation id. 

The entity embeddings are obtained by the final layer of the RGCN.  The major difference among RGCN and KGE is that RGCN embeddings are learned with graph convolutions and take into account relation features whereas the KGE embeddings are just supervised by equation \eqref{eq:kgeloss2}~\cite{schlichtkrull2018modeling,zheng2020dgl}.  Recaping, the RGCN relation embeddings combine both the graph structure information as well as the relation features generated by the semantic embedding methods. We implement and train the RGCN model using the DGL framework~\cite{wang2019deep}. The RGCN model was parametrized with $400$ hidden units per layer,  $L=2$ hidden layers. 

\subsection{Similarity Engine Construction}
\label{Similarity Engine Construction}
Our document similarity engine uses a combinations of the semantic and KGE embeddings as the RGCN model under-performed in certain ways as shown in Section \ref{Analysis}. Thereby we capture semantic information contained within a publication with the paper's topological information from the CKG e.g. papers, authors, concepts, topics, etc. relations.  Given a paper, the engine retrieves a list of top-k most similar papers using cosine distance. 

\section{Analysis}
\label{Analysis}

This section is organized into two parts presenting metrics and results evaluating the work done in Sections \ref{Structure of Graph} and \ref{Similar Paper Engine} respectively. Part one validates the construction and curation of the CKG by showing article topics align with common subject focuses of scientific journals and CKG relations are high quality. Part two analyzes the results of the similarity engine demonstrating we can improve the quality of recommended articles using both semantic and topological information. 

\subsection{Graph Validation}
\label{Graph Validation}

\subsubsection{Topic Model Validation}
\begin{figure}[t]
    \centering
    \includegraphics[width=\linewidth]{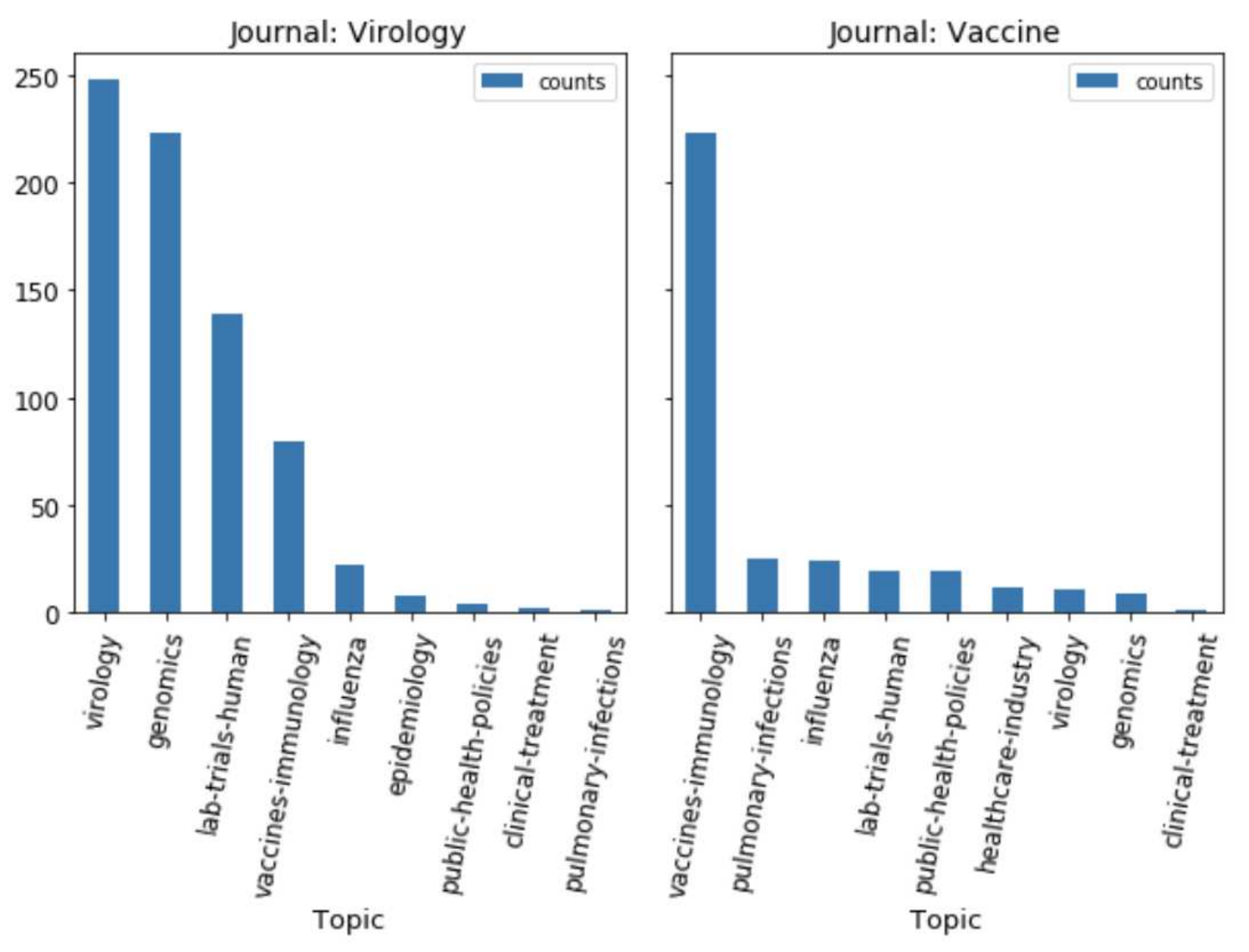}
    \caption{Distribution of topics by journal}
    \label{fig:topic_journal}
\end{figure}

\hfill\\
Most journals have well defined topics. For example, Journal of Virology explores the nature of viruses and mainly focuses on related domains; Journal of Vaccine focuses on the field of vaccinology. To evaluate our topic model, we summarise the generated topics from papers in the CKG belonging to these two journals in Figure~\ref{fig:topic_journal}. It can be seen that the generated topics of papers from Journal of Virology, e.g., virology, genomics and lab-trials-human, are highly related to virology. The generated topics of papers from Journal of Vaccine, e.g., vaccines-immunology, are highly related to vaccinology.

\begin{figure}[t]
    \centering
    \includegraphics[width=\linewidth]{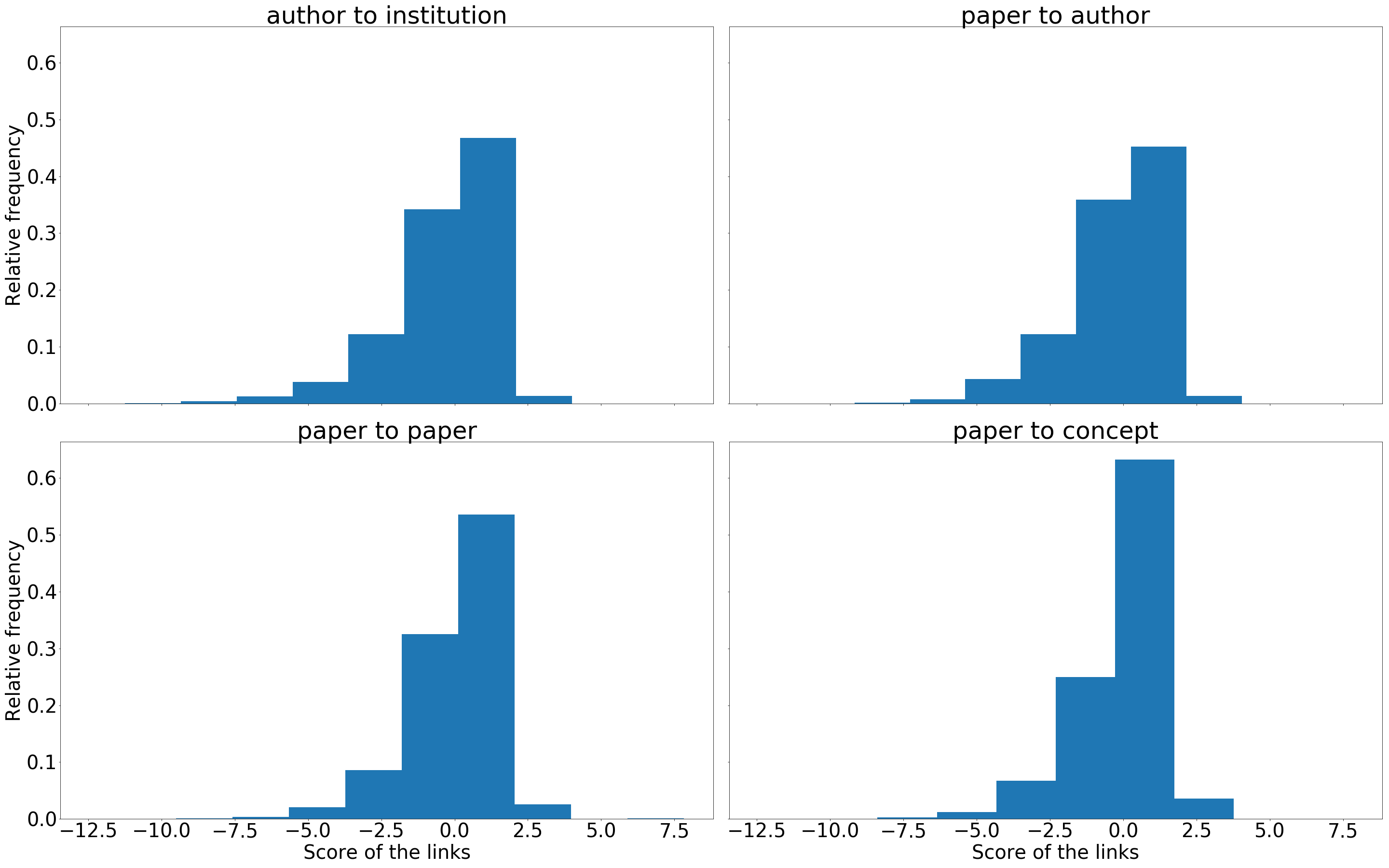}
    \caption{link prediction score distribution by relation types}
    \label{fig:link_prediction}
\end{figure}

\subsubsection{CKG Relation Validation}\hfill\\
To assess the correctness of the triplets that make up the CKG, we used the KGE model described in Section~\ref{TransE} to score each of its triplets using 
\begin{equation}\label{eq:kgescore}
    \mbox{\emph{score}} = \gamma - || h + r - t||_2,
\end{equation}
where $h$ and $t$ are the embeddings for the head and tail entities, $r$ is the embedding of the relation type, and $\gamma=12$ is an offset used to accelerate the training. We compute these scores for all of CKG's triplets by following a 10-fold strategy to split the triplets into 10 sets. In this approach, for each fold we used the remaining 9 folds to estimate the KGE model and used it to computed the scores for the left-out fold. According to Equation~\ref{eq:kgeloss2}, if the score computed for a triplet is around~0, then the triplet is consistent with the KGE model. On the other hand, if the score is further away from 0 (in either direction), then the triplet corresponds to potentially an outlier or an error. Figure~\ref{fig:link_prediction} shows the score distribution of the triplets for different relation types. These results show that the score of most triplets is close to 0 and that there is only a small fraction of inconsistent (according to the model) triplets.

\subsection{Recommendation Analysis}
\label{Recommendation Analysis}

\subsubsection{Topic Similarity}\hfill\\
We start by analyzing the topic similarity between each source paper and its \textit{top-5} most similar papers. In Table \ref{tab:topicSimilarity} a baseline is established by generating a \textit{top-5} list of papers random selected from the ~$42k$ scientific articles. Then, we collect \textit{top-5} similar article recommendations $r_{ij}$, $j<5$ for every source paper $s_i$ using four different embedding methods (Semantic, KGE, RGCN and Semantic\&KGE). We make use of topic-based distances to compute measures of similarity by creating a one-hot encoded vector $T(u)$ for every paper $p$ in our dataset representing its topics e.g. contains or not. Jaccard distance \cite{DBLP:journals/corr/Kosub16} is used to compute distance between vectors $u,v\in[T,F]^N, N\in\mathbb{N}$ 
\begin{align}\label{eq:jaccard}
J(u,v) = \frac{c_{TF}+c_{FT}}{c_{TT} + c_{TF} + c_{TF}}
\end{align}

\noindent where $c_{ij}$ is the number of occurrences of $u[k]=i$ and $v[k]=j$, $j<N$.\newline

\noindent Intra-List Similarity (ILS) \cite{ziegler2005improving} is used to measure topic similarity of paper recommendations using the average Jaccard distance between a source paper and its list of $top-5$ similar papers. We then take the average of scores over all source papers and compare across methods as displayed in Table \ref{tab:topicSimilarity}. For each source paper $s_i$ we define its topic similarity
\begin{align}\label{eq:ils1}
    TS(s_i) = \frac{1}{k=5}\sum^k_{j=1}J(T(s_i),T(r_{ij})),
    \quad
    TS = \frac{1}{N}\sum^N_{i=1}TS(s_i)
\end{align}
According to Equation~\ref{eq:ils1}, the lower the score, the more common topics are between the source paper and its $top-5$ similar papers.

\noindent In Table \ref{tab:topicSimilarity} we observe lower average Jaccard scores between source papers and similar recommendations relative to the baseline in all embedding methods. Furthermore, we note KGE embedding achieves a comparatively lower score than RGCN. Finally, the combination of semantic and KGE embeddings achieves the lowest Jaccard score. 

\vspace{-.3cm}
\begin{table}[H]
\caption{Topic similarity (Jaccard distance) of recommendations vs random baseline.}
    \begin{tabular}{lc}\toprule
    \textbf{Method}      & \textbf{Topic Similarity$_{Jaccard}$} \\ \midrule
    Random               & .821  \\
    Semantic$_{Sem}$     & .360  \\ 
    Graph$_{KGE}$        & .345  \\ 
    Graph$_{RGCN}$       & .654  \\ 
    \textit{Sem. \& KGE} & .311  \\ \bottomrule
    \end{tabular}
\centering
\label{tab:topicSimilarity}
\end{table}
\vspace{-.2cm}

\subsubsection{Citation Similarity}\hfill\\

The CKG citation network shows the relationship between papers. If a paper is cited by another, they may share the same topic, use similar technology or have similar motivation. We train RGCN embeddings from the CKG with and without the citation network and follow the same methodology for KGE embeddings. We select only papers that cite at least one other paper. For each of these papers we generate the \textit{top-5} similar papers and calculate the average number of a paper's citations that appear in the \textit{top-5} recommended most similar papers. For Table \ref{tab:citations} we average this score across all papers for the four RGCN and KGE embeddings. We observe that KGE trained with citations has the highest overlap score at 29.11\% as expected. Further, KGE embeddings learned without citations do a poor job of recommending cited papers in the \textit{top-5}. This is expected since the relations \textit{authored\_by}, \textit{associated\_topic}, and \textit{associated\_concept} do not give much information to infer the exact citation: many papers share the same topic and concept. 

\vspace{-0.2cm}
\begin{table}[H]
    \caption{RGCN vs KGE Top-5 Citation Overlap}
    \begin{tabular}{l|c}\toprule
      \textbf{Method} & \textbf{Overlap}\\ \midrule
      \textbf{\textit{RGCN$_{without\ citations}$}}   & 5.22\%  \\
      \textbf{\textit{KGE$_{without\ citations}$}}   & 0.01\%  \\
      \textbf{\textit{RGCN$_{with\ citations}$}}   & 8.96\%  \\
      \textbf{\textit{KGE$_{with\ citations}$}}    & 29.11\%  \\
    \end{tabular}
    \centering
    \label{tab:citations}
\end{table}

\begin{figure}[t]
    \captionsetup{format=plain}
    \centering
    \includegraphics[width=\linewidth]{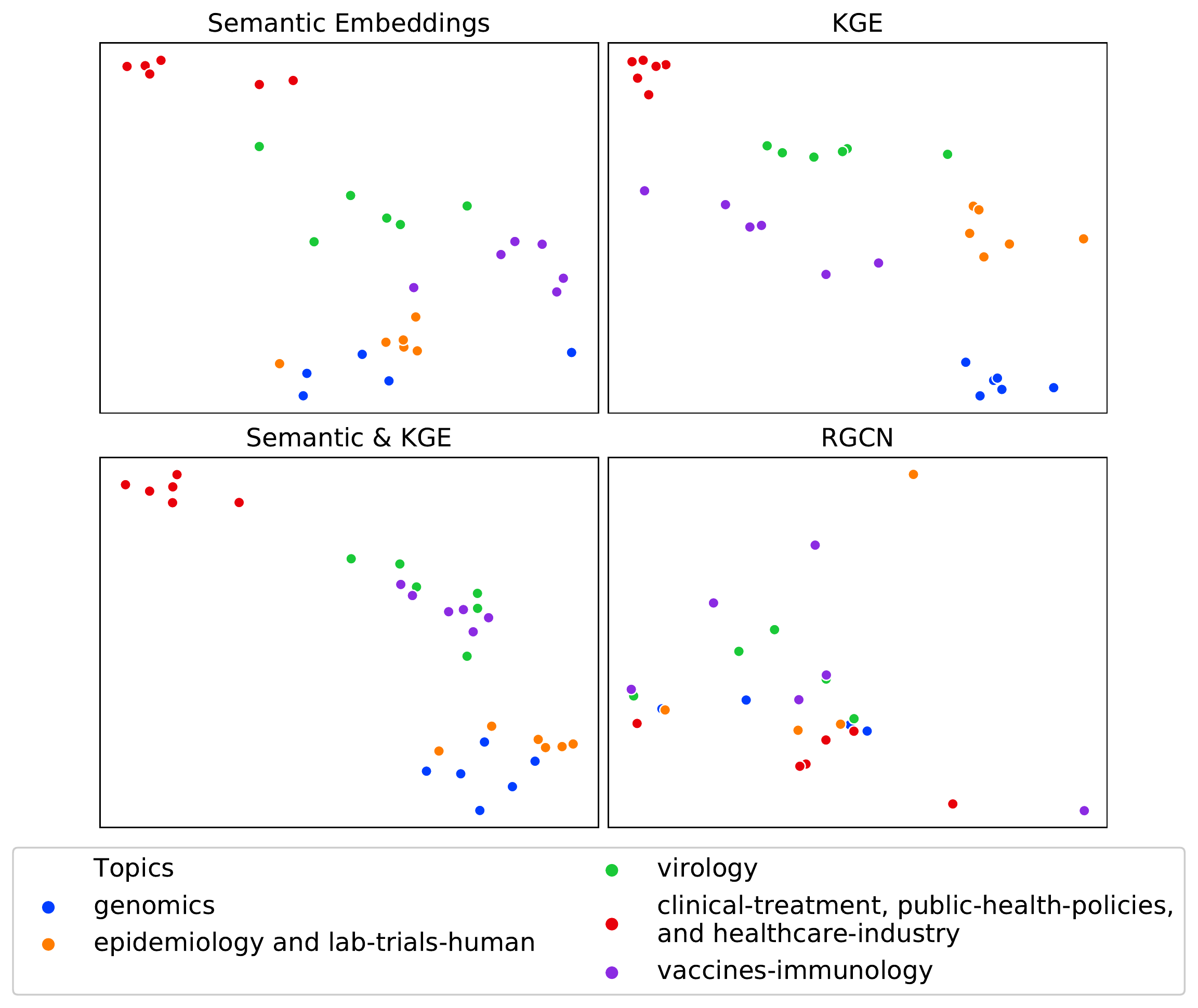}
    \caption{Visual comparison of truncated SVD of four embedding methods using five scientific articles in the dataset. Paper CORD\_UIDs: \textcolor{blue}{pw60qx7c}, \textcolor{orange}{fjfc3rto}, \textcolor{green}{790d7v7q}, \textcolor{red}{v2lp739t}, \textcolor{mypurple}{kt5awf8i}}
    \label{fig:truncatedSVD}
\end{figure}
\vspace{-.8cm}

\vspace{.2cm}
\begin{table}[H]
    \caption{Overlapping (intersection over union) scores of Top-5 similar papers by methodology}
    \begin{tabular}{l|cccc}\toprule
        & \textbf{Random}  & \textbf{Semantic$_{Sem}$}  & \textbf{\textit{KGE}} & \textbf{\textit{RGCN}} \\ \midrule
    \textbf{Random}          & 1.000 & 0.014 & 0.009    & 0.008 \\ 
    \textbf{Semantic$_{Sem}$}& -     & 1.000 & 0.084    & 0.081 \\ 
    \textbf{Graph$_{KGE}$}   & -     & -     & 1.000    & 0.137 \\ 
    \textbf{Graph$_{RGCN}$}  & -     & -     & -        & 1.000 \\ \midrule
    \textit{Sem \& KGE}      & 0.10  & 0.164 & 0.463    & 0.005 \\ \bottomrule
    \end{tabular}
    \centering
    \label{tab:overlapsets}
\end{table}

\subsubsection{Embedding Subspace}\hfill\\
We use truncated singular value decomposition (SVD) to create 2D projections of paper embeddings of different embeddings methods. We select 5 papers with different topics in our dataset and their corresponding \textit{top-5} recommendations. We plot the truncated SVD reduction of their embeddings and plot them based on the source paper. The results are represented in Figure \ref{fig:truncatedSVD}. The top left shows the SciBERT embeddings for the five papers and their associated topics (color coded according to paper). We observe the topics \textit{genomics} and \textit{epidemiology and human lab trials} as described in Section \ref{Structure of Graph}, are close to each other. This is expected as many genomic studies are genome wide association studies, which are considered a subset of epidemiology. 
The top right shows the KGE embeddings' SVD result. It can be seen papers from same topics are clustering to each other while separating across topics.
On the other side, the combination of SciBERT embeddings with KGE embeddings which is currently used in the similarity engine (bottom left) shows that 
\textit{virology} and \textit{vaccines immunology}, and \textit{genomics} and \textit{epidemiology and human lab trials} narrow in proximity from KGE. This matches the observed research given \textit{virology} is the study of viruses while similarly, \textit{vaccines immunology} is the study of how viral immunizations stimulate the immune system hence closer embedding similarity match expectations of researchers. \newline

\begin{figure}[t]
    \captionsetup{format=plain}
    \includegraphics[width=\linewidth]{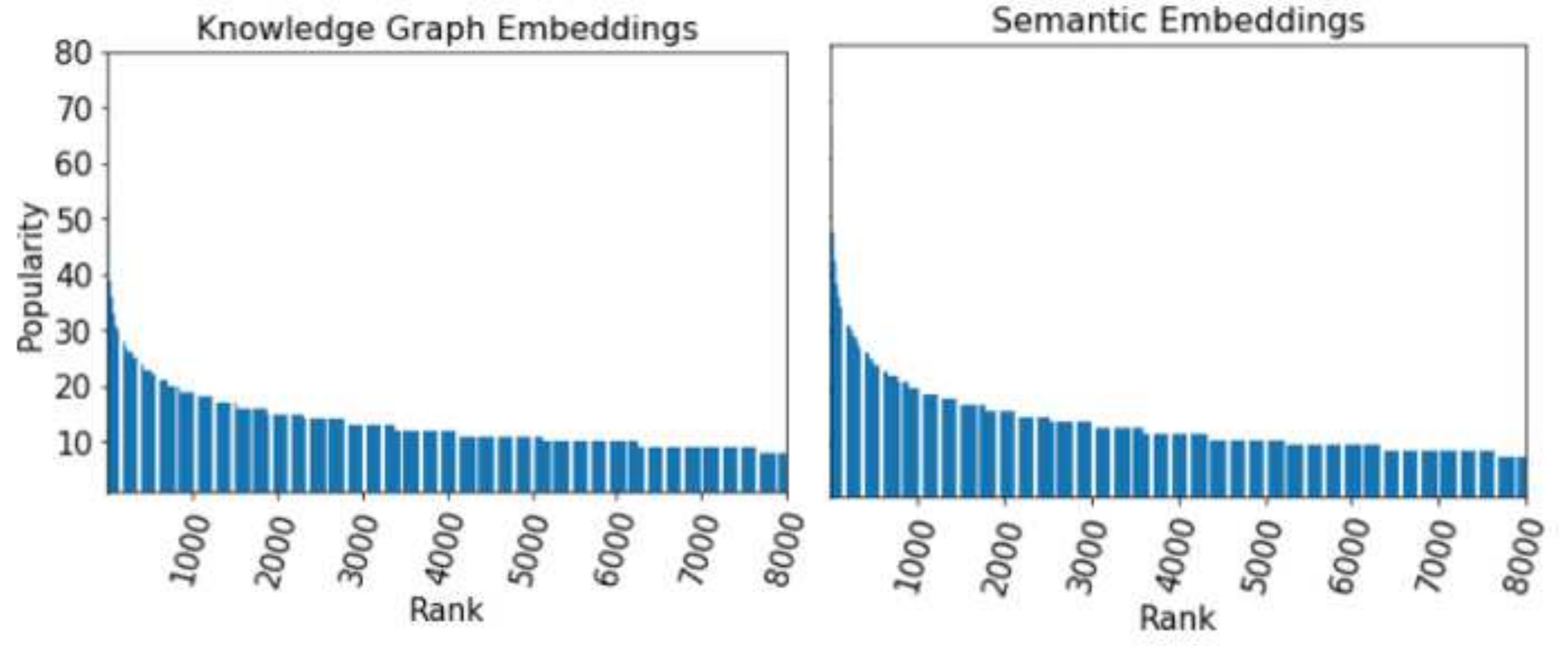}
    \caption{Popularity (= occurrences of paper in the top-5 most similar paper list) analysis for semantic embedding and KGE embedding engine grouped by bins.}
    \label{fig:popanalysis}
    \centering
\end{figure}
\vspace{-0.3cm}

\subsubsection{Recommendation Overlap}\hfill\\
We generate \textit{top-5} most similar papers for each paper in the dataset using five different methodologies, Random (Randomly select 5 papers), Semantic, KGE, RGCN and Semantic\&KGE. Table \ref{tab:overlapsets} captures the intersection over union of similar paper sets across methodologies. We observe a low overlapping between semantic and graph embeddings, which is as expected since Semantic capture the semantic information of certain paper while KGE/RGCN capture the topological information of the CKG. The combination of them, i.e. Semantic\&KGE, shows the agreement with both side, which means it can recommend papers with a conjunction of both semantic and topological information..

\subsubsection{Popularity}\hfill\\
Figure \ref{fig:popanalysis} presents a popularity analysis of KGE and Semantic Embedding, where popularity captures the number of occurrences of an individual paper in the \textit{top-5} most similar items list for all papers in the dataset grouped by frequency. The left tail of the distribution shows papers that occur many time times in \textit{top-5} recommended lists with the overall distribution resembling a power law distribution common to recommendation systems~\cite{jannach2013recommenders}. For KGE embeddings 707 papers occur more than 20 times and for semantic 912 occur more than 20 times.

\section{Conclusion}

In this paper we construct a COVID-19 Knowledge Graph from the CORD-19 dataset and demonstrate how researchers and policy makers can extract timely information to answer key scientific questions on COVID-19 from a corpus of scientific articles. To further facilitate efforts we employ  machine learning entity detection models to extract medical entities and relationships. 
With the help of medical professionals we add global topic information that forms additional medical relationships in the CKG. We train KGE models using CKG relations to obtain paper embeddings capturing topological isomorphic and semantic information for the application of similar paper retrieval on \url{www.cord19.aws}. Future work may include further enhancements to CKG information retrieval capabilities such as: expanding biomedical entity extraction using biomedical concept annotators like PubTator\footnote{https://www.ncbi.nlm.nih.gov/research/pubtator/}, re-training RGCN models with additional entity and relation attributes, and incorporating additional KGs into the CKG e.g. COVID-19 drug repurposing graphs~\cite{gramatica2014graph}.

\bibliographystyle{ACM-Reference-Format}
\bibliography{cikm2020}

\end{document}